\documentclass[prl,twocolumn,showpacs,floatfix,amsmath,nofootinbib,amssymb,floatfix]{revtex4}
\usepackage{graphicx,color,dcolumn,booktabs,bm}
\usepackage{longtable,lscape}
\usepackage{txfonts}
\usepackage{overpic}
\usepackage{amssymb}
\usepackage{indentfirst}
\usepackage{feynmf}   
\usepackage{slashed}  
\usepackage{cases}
\usepackage{color}
\usepackage{multirow}
\usepackage{epstopdf}
\usepackage{float}
\usepackage{graphicx,color,dcolumn,booktabs,bm}
\usepackage[colorlinks, citecolor=blue,anchorcolor=red,menucolor=red, linkcolor=red,filecolor=red,runcolor=red,urlcolor=blue,frenchlinks=red]{hyperref}

\graphicspath{{Figures/}} %

\begin{document}

\title{Predicting New Above-Threshold Molecular States Via Triangular Singularities}

\author{Yin Huang$^{1}$}\email{huangy2019@swjtu.edu.cn}
\author{Xurong Chen$^{2,3}$} \email{xchen@impcas.ac.cn}
\affiliation{$^{1}$School of Physical Science and Technology, Southwest Jiaotong University, Chengdu 610031,China}
\affiliation{$^{2}$Institute of Modern Physics, Chinese Academy of Sciences, Lanzhou 730000, China }
\affiliation{$^{3}$School of Nuclear Science and Technology, University of Chinese Academy of Sciences, Beijing 100049, China}

\begin{abstract}
Considering that the experimentally observed molecular states are significantly fewer than those predicted theoretically, and that these states are traditionally classified as lying below thresholds while several candidates are found above them, we propose to broaden the definition of molecular states to include those that exist just above the thresholds.
Identifying resonance peaks in invariant mass distributions and scattering cross-sections is crucial for probing these states, yet the mechanisms responsible for such enhancements remain unclear, complicating our understanding of new particle production.  While the peaks linked to triangular singularities do not correspond to true hadronic states,
the associated production mechanisms may provide valuable insight into the search for genuine hadrons. In this work, we propose employing the triangular singularity mechanism to theoretically investigate yet-to-be-observed molecular states, particularly those that could test heavy quark symmetry. We argue that these states may have true masses surpassing the thresholds of their constituent components, rather than being predicted to be below them by theoretical models. Our findings suggest the possible existence of 18 predicted heavy quark molecular states, including $X(4014)$, $Z_{cs}(4123)$, $X_{c0}(4500)$, $X_{c1}(4685)$, $Y(4320)$, $Z(4430)$, and
$\Upsilon(11020)$, which are posited to contain $D^{*}\bar{D}^{*}$, $D^{*}\bar{D}^{*}_{s}$, \( D_{s1}^{+}D_{s}^{-} \),\( D_{s1}^{+}D_{s}^{*-} \), $D_1\bar{D}$, $D_1\bar{D}^{*}$, and $B_{1}(5721)\bar{B}$ constituents, respectively.
The recognition of these states would substantiate heavy quark symmetry and enhance our understanding of hadronic dynamics and molecular states formation.
\end{abstract}

\maketitle
\textit{Introduction}--Heavy quark symmetry (HQS)~\cite{Isgur:1991wq, Liu:2019zoy} is a key theoretical tool in molecular state research, but many of its predicted states remain unobserved,
raising doubts about its validity. For example, heavy quark flavor symmetry (HQFS) predicts hidden-bottom pentaquark molecules, based on several hidden-charm candidates like \(P_c(4312)\),
\(P_c(4380)\), \(P_c(4440)\), \(P_c(4457)\), \(P_{cs}(4338)\), and \(P_{cs}(4459)\)~\cite{LHCb:2015yax, LHCb:2016ztz, LHCb:2016lve, LHCb:2019kea, LHCb:2020jpq, LHCb:2022ogu, Chen:2019bip,
Guo:2019fdo, Xiao:2019aya, He:2019ify, Xiao:2019mvs, Roca:2015dva, Chen:2015moa, Chen:2015loa, Yang:2015bmv, Huang:2015uda, Du:2019pij}. However, no hidden-bottom counterparts have been observed.
Additionally, heavy quark spin symmetry (HQSS) predicts a partner state of \(X(3872)\) with \(J^{PC} = 2^{++}\) and a \(D^{*}\bar{D}^{*}\) molecular component~\cite{Guo:2013sya}. While the Belle
Collaboration suggested its existence in 2022 via \(\gamma\gamma \to \gamma\psi(2S)\), the global significance is only \(2.8~\sigma\)~\cite{Belle:2021nuv}, highlighting the need for further
investigation.

We assert the validity of HQS, acknowledging that any challenges to this framework necessitate rigorous theoretical and experimental scrutiny.
Our main point is that the molecular states theoretically predicted to probe HQS do indeed exist; however, their experimental detection is
challenging primarily because these molecular states exist above the threshold, rather than below it as typically anticipated. In this Letter, we propose using
the triangular singularity mechanism to search for these above-threshold molecular states.  This effort is guided by existing data on molecular states, particularly
those pertinent for validating the accuracy of HQS.

\begin{figure}[http]
\begin{center}
\includegraphics[bb=50 560 1050 710, clip, scale=0.42]{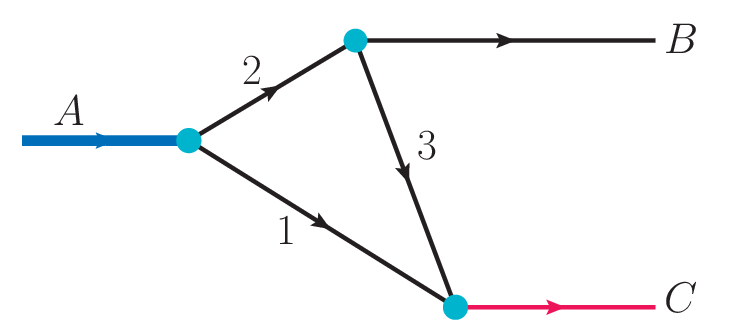}
\caption{The Feynman diagrams for the decay of \( A \) to \( BC \) via a triangle loop involve intermediate states 1, 2, and 3, where \( A \) is the heavy-quark hadron molecule to identify, and \( C \) is a known molecular state with components 1 and 3.}\label{cc1}
\end{center}
\end{figure}

Triangular singularities (TS), initially introduced by Landau~\cite{Karplus:1958zz, Landau:1959fi}, are characterized by pronounced observable peaks in the invariant mass distribution or scattering cross-section of specific reactions, despite not representing actual physical states. So far, many related physical phenomena have been observed,
with a notable example being the significant isospin violations in the $\eta(1405/1475) \to \pi\pi\pi$ reaction~\cite{BESIII:2012aa}, which can be naturally explained by
TS~\cite{ Wu:2011yx}.  In contrast to the ambiguous mechanisms underlying the enhancements observed in resonant states, the production mechanism of TS peaks is well-characterized.  A comprehensive discussion of the details can be found in Refs.~\cite{Guo:2017jvc,Guo:2015umn,Guo:2010ak,Guo:2019twa} and will not be elaborated upon here.  We only present its Feynman diagram in Fig.~\ref{cc1} and note that, according to the Coleman-Norton theorem~\cite{Coleman:1965xm},  the existence of TS depends on whether the decay process the classical nature of the decay process and the simultaneous on-shell and collinear conditions of all three internal particles in the rest frame of the decaying particle.

Intriguingly, recent work outlined in Ref.~\cite{Sakthivasan:2024uwd} indicates that strong re-scattering between the final-state particles 1 and 3 does not influence the position of the triangular singularity or its characteristic linear structure in the invariant mass spectrum $M_{13}$. Consequently, when particles 1 and 3 are interpreted as molecular components of hadron $C$, the re-scattering event leading to the formation of $C$, the TS signal provides direct experimental
evidence for the molecular structure of $C$.  Subsequently, we will consider the molecular state $C$ as a known particle and eelucidate how to utilize triangular
singularities to probe for additional molecular state structures.

\textit{Our Strategies}--As illustrated in Fig.~\ref{cc1}, the formation of a TS necessitates an initial hadron, two
final state hadrons, and three intermediate particles.  Assuming that hadron $C$ is a known molecular state, experimentally identified and
composed of intermediate particles 1 and 3, and that particle 2 can be effectively associated with the external hadron $B$ through the
reaction $2 \to B + 3$, the principal challenge resides in validating the existence of the initial hadron $A$.   Confirming the presence of hadron
\(A\)  is essential to complete the triangular singularity framework, thereby yielding a kinematic configuration that is amenable to experimental verification.

For simplicity, we treat hadron \( A \) as an unknown particle to be predicted. A peak arises if the masses and four-momenta of the involved particles satisfy the Landau equation~\cite{Landau:1959fi}:
\begin{align}
1 + 2y_{12}y_{23}y_{13} = y_{12}^2 + y_{23}^2 + y_{13}^2,\label{eq1}
\end{align}
where \( y_{ij} = (m_i^2 + m_j^2 - p_{ij}^2)/(2m_i m_j) \), with \( m_i \) \((i=1,2,3)\) representing the masses of intermediate particles, and \( p_{ij}^2 = (p_i + p_j)^2 = m_{ij}^2 \) being the four-momentum of the \( ij \) pair, which corresponds to the external particle. Solving Eq.~\ref{eq1} yields the mass of the initial state particle \( A \):
\begin{align}
M^2 = m_1^2 + m_2^2 + 2E_1E_2 \pm 2|\vec{p}_{1cm}||\vec{p}_{2cm}|,\label{eq2}
\end{align}
where \( E_1, E_2 \) and \( \vec{p}_{1cm}, \vec{p}_{2cm} \) denote the energies and three-momenta of these particles in the rest frame of intermediate particle 3.
Among the two solutions for the mass (Eq.~\ref{eq2}), only the lower root lies within the allowed region. Detailed discussions and examples are given in Refs.~\cite{Schmid:1967ojm} and~\cite{Guo:2015umn}. From Eq.~\ref{eq2}, the permitted mass range for particle \( A \) is determined by the triangular singularity within the allowed domain:
\begin{align}
M^2 \in \left[(m_1 + m_2)^2, m_1^2 + m_2^2 + 2m_1 \frac{m_3^2 + m_2^2 - m_B^2}{2m_3} \right].\label{eq3}
\end{align}
\( M \) is minimum when particles 1 and 2 are at rest, and maximum when particles 1 and 3 are at rest. Using these values in Eq.~\ref{eq1}, the mass range for particle \( C \) is:
\begin{align}
m_C^2 \in \left[(m_1 + m_3)^2, m_1^2 + m_3^2 + 2m_1 \frac{m_2^2 + m_3^2 - m_B^2}{2m_2} \right]. \label{eq4}
\end{align}

To accurately measure the mass of particle \( A \), the discrepancy between the maximum and minimum values should be small, as given by:
\begin{align}
\delta M^2= \frac{m_1}{m_3} \delta(23B) \left[\delta(23B) + 2m_B\right], \label{eq5}
\end{align}
where \( \delta(23B) = (m_2 - m_3 - m_B) \). This suggests that for precise mass determination, \( m_2 \) should be close to the decay product thresholds,
\( m_3 \) and \( m_B \). Detailed examples can be found in Ref.~\cite{Liu:2015taa}. This condition also implies that the mass of the hadronic molecular state
\( C \) is near the thresholds of its components (see Eq.~\ref{eq4}). In fact, molecular states are typically close to these component thresholds, and if \( m_B \)
is small at same time, such as for \( \gamma \) or \( \pi \), it further aids in the accurate measurement of particle \( A \)'s mass. The small variation in \( M \) indicates
that \( M \) is near the thresholds of particles 1 and 2, allowing us to treat \( A \) as a bound state of particles 1 and 2.

As shown in Eqs.~\ref{eq3}-\ref{eq4}, to observe a TS signal experimentally, the mass of particle \(C\) (or \(A\)) must be at least as large as the thresholds of its components. Currently, experimentally observed particles containing heavy quarks, which can be interpreted as molecular states with masses exceeding their thresholds, include \( T^{++}_{c\bar{s}0}(2900) \), \( X(3872) \), \( Z_{cs}(3985) \), \( X(4020/4025) \), \( Z_b(10610) \), \( Z_b(10650) \), and \( P_{cs}(4338) \). Details on these hadrons, including molecular assignments and masses, are given in Tab.~\ref{tab-1}. This provides an opportunity to use TS to search for molecular states $A$ with heavy quarks. Observing $A$ could help confirm the molecular structure of hadron \( C \). Since the mass of \( T^{++}_{c\bar{s}0}(2900) \) exceeds its molecular threshold by about 20 MeV, we search for a molecule \( A \) whose mass also exceeds its threshold by around 20 MeV.
\begin{table}[http!]
\centering
\caption{Experimental states (first column) that could be interpreted as molecular states (third column), with masses (second column)
greater than the threshold (last column) of the molecular components. Note that the particle masses we use are all taken as central
values~\cite{ParticleDataGroup:2022pth} and are given in units of MeV. }\label{tab-1}
\setlength{\tabcolsep}{-3mm}{
\begin{tabular}{ccccc}
\hline\hline
~~ States                 &~~~~~~~~~ Exp.values              &~~~~~~~~~~~~~            Component               ~~~   &~~~~~~~~~ Threshold             ~~     \\ \hline
~~ $X(3872)^0$            &~~~~~~~~~ $3871.69$               &~~~~~~~~~~~~~ $D^{*-}D^{+}/\bar{D}^{*0}D^0$      ~~~   &~~~~~~~~~ $3879.92/3871.69$     ~~     \\
~~ $X(4020)^0$            &~~~~~~~~~ $4022.9 $               &~~~~~~~~~~~~~ $D^{*-}D^{*+}/\bar{D}^{*0}D^{*0}$  ~~~   &~~~~~~~~~ $4020.52/4013.70$     ~~     \\
~~ $X(4025)^0$            &~~~~~~~~~ $4026.30$               &~~~~~~~~~~~~~ $D^{*-}D^{*+}/\bar{D}^{*0}D^{*0}$  ~~~   &~~~~~~~~~ $4020.52/4013.70$     ~~     \\
~~ $Z_b(10610)^{0}$       &~~~~~~~~~ $10609.0$                 &~~~~~~~~~~~~~ $B^{+}B^{*-}/B^{0}\bar{B}^{*0}$    ~~~   &~~~~~~~~~ $10604.16/10605.38$   ~~   \\
~~ $Z_b(10650)^{+}$       &~~~~~~~~~ $10652.2$                 &~~~~~~~~~~~~~ $B^{*+}\bar{B}^{*0}$  ~~~                 &~~~~~~~~~ $10650.41$   ~~    \\
~~ $P_{cs}(4338)^0$       &~~~~~~~~~ $4338.2$                 &~~~~~~~~~~~~~ $\Xi_c^{+}D^{-}/\Xi_c^{0}\bar{D}^{0}$  ~~~   &~~~~~~~~~ $4337.37/4335.28$   ~~  \\
~~ $Z_{cs}(3985)^0$       &~~~~~~~~~ $3982.5$                 &~~~~~~~~~~~~~ $D_s^{*-}D^{+}/D_s^{-}D^{*+}$  ~~~   &~~~~~~~~~ $3981.8/3978.61$   ~~           \\
~~ $T^{++}_{c\bar{s}0}(2900)$ &~~~~~~~~~ $2921.0$              &~~~~~~~~~~~~~ $K^{*+}D^{*+}$  ~~~   &~~~~~~~~~ $2901.93$   ~~           \\
\hline \hline
\end{tabular}}
\end{table}

In this analysis, we posit that molecular states $A$ and $C$ serve as critical examples for demonstrating how triangular singularities
can be employed to validate heavy quark symmetry. The decay process illustrated in the triangular diagram reveals that the molecular
states $A$ and $C$ share a common heavy-quark-containing component (particle 1). The additional
components, particles 2 and 3, interact via the decay channel $2\to{}3+B$.  When particles 2 and 3 form a pair of heavy quark partner states,
 it creates a unique opportunity to leverage triangular singularities as a means of identifying states that conform to the conditions of
 heavy quark partners. Upon successful experimental validation of these states, they can serve as a robust framework for testing heavy
 quark symmetry. This approach not only enhances our understanding of the underlying principles of heavy quarks but also offers a pathway
 to exploring novel states within the molecular landscape.

\textit{Results and Discussions}--
\textbf{1.~Hidden-Charm Tetraquark Molecules:}
First, we consider the final state particle $C$ as $X(3872)$, with molecular components $\bar{D}^{*}$ and $D$. Given $m_C = 3871.69$ MeV, only $\bar{D}^{*0}D^0$
meets the condition \(m_C \geq m_1 + m_3\) (see Tab.~\ref{tab-1}). To minimize \(\delta(23B)\), we select \(D^{*}\), \(D^0\), and \(\pi\) for particles 2, 3,
and $B$. This gives $\delta(D^{*\pm}D^0\pi^{\pm}, D^{*0}D^0\pi^0) = (5.85, 7.03)
\text{ MeV}$, yielding $M(A)$ as \(4017.11-4017.33\) MeV for \(\delta(D^{*\pm}D^0\pi^{\pm})\) and \(4013.70-4013.96\) MeV for
\(\delta(D^{*0}D^0\pi^0)\). This state, sharing $D^{*}$ and $D$ as HQSS partners, may be a partner of $X(3872)$ but lacks experimental confirmation, with only
slight evidence from \(e^{+}e^{-} \to D^{*}\bar{D}\)~\cite{Dubynskiy:2006sg, CLEO:2008ojp}.

In 2022, the Belle Collaboration reported cross-section measurements for the two-photon process \(\gamma\gamma \to \gamma\psi(2S)\) up to 4.2 GeV, identifying a
new state with mass \(M = (4014.3 \pm 4.0 \pm 1.5)\) MeV and width \(\Gamma = 4 \pm 11 \pm 6\) MeV~\cite{Belle:2021nuv}. This particle is considered a \(D^{*}\bar{D}^{*}\)
molecular state with \(J^{pc} = 2^{++}\), potentially a partner of \(X(3872)\). However, its global significance is limited to 2.8 $\sigma$, requiring further experimental
validation. Replacing the \(\pi\) meson with a photon in the reaction \(A \to D^{*0} \bar{D}^{*0} \to \pi^0 \bar{D}^{0} D^{*0}\) gives \(\delta(D^{*0} D^{0} \gamma)
\approx 142.0\) MeV, corresponding to a mass range of \(M = 4013.7 - 4016.4\) MeV, consistent with Belle$'$s result.  Guo used this process to precisely measure the mass of
$X(3872)$~\cite{Guo:2019qcn}, observing a significant peak in the $X(3872)\gamma$ production within the range of initial hadron mass variations (see Fig.~2 in Ref.~\cite{Guo:2019qcn}).
Subsequent experiments confirmed the presence of $X(3872)$ in this process~\cite{BESIII:2023hml}, though they provided the invariant mass spectrum of \( D^0 \bar{D}^0 \pi^0 \),
not \( \gamma X(3872) \). This sharp peak remains indirect evidence of its existence.

We can choose intermediate particles 1 and 3 as \(D^0\) and \(D^{*0}\), with particle 2 identified as \(D_1(2420)\), which decays predominantly to \(D^{*}\pi\)~\cite{ParticleDataGroup:2022pth},
implying the final state particle \(B\) is a pion.  This primary decay mode will produce a sharper peak in the production line shape. Our calculations show that to generate a triangular singularity
peak, the initial state mass must be within \(M = 4286.94 - 4303.62 \, \text{MeV}\) or \(M = 4290.94 - 4307.83 \, \text{MeV}\). This particle, denoted
\( Y(4320)\), was observed by BESIII in the \( e^{+}e^{-} \to \pi^{+}\pi^{-} J/\psi \) channel~\cite{BESIII:2022qal} with measured values \( M = 4298 \pm 12
\pm 26 \, \text{MeV}\) and \( \Gamma = 127 \pm 17 \pm 10 \, \text{MeV}\). Notably, this may represent a high partial-wave molecular state of \(D_1 \bar{D}\),
as the \(S\)-wave molecular state with a \(D_1 \bar{D}\) component~\cite{Ji:2022blw,Wang:2013cya} is typically associated with the experimentally observed \(Y(4230)\)~\cite{BESIII:2018iea}.

Similar to \( Y(4230) \) and its partner \( Y(4360) \), \( Y(4320) \) also has a partner, \( Z(4430) \). We consider the final state particle \(C\) as \(X(4020)\) or \(X(4025)\), both interpreted as \(\bar{D}^{*}D^{*}\) molecules. In this setup, intermediate particles 1 and 3 are \(\bar{D}^{*}\) and \(D^{*}\), while particle 2 is taken as \(D_1(2420)\). The mass of the initial particle \(A\) is independent of whether \(C\) is \(X(4020)\) or \(X(4025)\), depending instead on molecular components and the intermediate particle 2. Our calculations show that for \(C = X(4020)\) with \(\delta(D_1(2420)^{\pm}D^{*\pm}\pi^0) = 280.86\) MeV, \(A\) has a mass range of \(4436.36 - 4453.76\) MeV. When \(\delta(D_1(2420)^{0}D^{*0}\pi^0) = 280.27\) MeV, the mass range is \(4428.95 - 4446.33\) MeV. A charged partner may also exist with mass ranges between \(4432.36 - 4450.54\) MeV.

Currently, only one charged charmonium-like state, \( Z(4430) \), has been reported in the predicted region, but its mass and width are not fully determined. The Particle Data Group provides an average mass of \( 4478^{+15}_{-18} \) MeV and a width of \( 181 \pm 31 \) MeV~\cite{ParticleDataGroup:2022pth}. Belle measurements report masses of \( 4433 \pm 4 \pm 2 \) MeV~\cite{Belle:2007hrb}, \( 4443^{+15+19}_{-12-13} \) MeV~\cite{Belle:2009lvn}, and \( 4485^{+22+28}_{-22-11} \) MeV~\cite{Belle:2013shl}, with corresponding widths. LHCb reports a mass of \( 4475 \pm 7^{+15}_{-25} \) MeV and a width of \( 172 \pm 13^{+37}_{-34} \) MeV~\cite{LHCb:2014zfx}. If \( Z(4430) \) is a \( D_1 \bar{D}^{*} \) molecular state, our method can accurately measure its mass.

\textbf{2.~Hidden-Charm tetraquark molecules with strange:}
Exotic states with \( c\bar{c}\bar{q}q \) quarks are expected to have \( SU(3) \) partners \( Z_{cs} \) containing \( c\bar{c}\bar{s}q \). So far, only two such tetraquark mesons have been observed~\cite{ParticleDataGroup:2022pth}, prompting further searches for \( Z_{cs} \) states. We propose searching for \( Z_{cs} \) states decaying into \( KX(4020) \) or \( KX(4025) \) via \( e^{+}e^{-} \to KKX(4020)/X(4025) \). Using the triangular singularity mechanism, we predict the mass of \( Z_{cs} \) to be \( M = 4545.37 - 4548.43 \) MeV for \( \delta(D_{s1}(2536)^{+}D^{*+}K^{0}) = 27.24 \) MeV. We choose \( D_{s1}(2536) \) due to its dominant \( K^0D^{*+} \) decay mode. Other \( D_{sJ} \) states are excluded due to unknown branching ratios or distance from the \( D^{*}K \) threshold. We also suggest searching for the partner state \( D_{s1}(2536)^{+}\bar{D}^{-} \) by replacing \( C \) with \( X(3872) \), with a predicted mass range of \( M = 4404.77 - 4407.71 \) MeV.

Experiments show that both the ground state \(D_s\) and first excited state \(D_s^*\) of \(D_{s1}(2536)\) interact with \(D^*\) and \(D\) to form the molecular state \(Z_{cs}(3985)\)~\cite{BESIII:2023wqy,BESIII:2020qkh}, motivating further searches for other \(Z_{cs}\) states. We treat \(Z_{cs}(3985)\) as particle \(C\), with \(D^{*+}\) and \(D_s^{-}\) as its components. To minimize \(\delta(23B)\), we set \(B\) as a photon and particle 2 as \(D_s^*\), decaying predominantly to \(\gamma D_s\). Under these conditions, a \(Z_{cs}\) molecular state with \(D^{*+}D_s^{*-}\) and mass \(M = 4122.46 - 4125.02\) MeV could exhibit a significant triangle singularity peak. This state could be the partner of \(Z_{cs}(3985)\). Replacing the photon with \(\pi^0\) yields a more precise mass range of \(M = 4122.46 - 4122.77\) MeV. The existence of a \(D^{*}\bar{D}_s^{*}\) molecular state around 4100 MeV with \(J^{PC} = 1^{+-}\) was predicted in Ref.~\cite{Wang:2020dgr}, and supported by recent BESIII data~\cite{BESIII:2022vxd}, which measured the mass as \(M = 4123.5 \pm 0.7 \pm 4.7\) MeV with a significance of 2.1 \(\sigma\).

The possibility of a bound state in the \( D^{(*)}DK \) system has garnered attention. Despite various predictions for the mass spectrum~\cite{SanchezSanchez:2017xtl, MartinezTorres:2018zbl, Wu:2019vsy} and decay properties~\cite{Huang:2019qmw}, experimental evidence remains lacking~\cite{Belle:2020xca}. We propose utilizing the triangle singularity mechanism to search for \( D^{*}DK \) three-body molecular states, treating the \( D_s^{-} \) and \( D^{*+} \) components of \( Z_{cs}(3985) \) as particles 3 and 1 in the triangle diagram. Particle 2 is identified as \( D_{s0}(2317) \), a molecular state formed by \( KD \), and \( B \) is \( \pi^0 \). This results in a mass range of \( M = 4328.06 - 4340.06 \) MeV for state \( A \). Replacing particles 3 and 1 with \( D_s^{*-} \) and \( D^{+} \), which are also components of \( Z_{cs}(3985) \), gives a mass range of \( M = 4329.16 - 4339.62 \) MeV, where \( B \) remains \( \pi^0 \) and particle 2 is replaced with \( D_{s1}(2460) \), a \( KD^{*} \)-formed molecular state, providing a more precise mass measurement.

We propose that experiments search for hidden-charm tetraquarks with \( s\bar{s} \) in the \( K^0Z_{cs}(3985) \) final state. Based on the triangular singularity mechanism, we predict its mass to be \( M = 4503.46{-}4506.49 \, \text{MeV} \), with a \( D_{s1}^{+}D_{s}^{-} \) molecular component. Recently, the LHCb collaboration observed such a particle in the \( J/\psi\phi \) spectrum, with a mass of \( M = 4503.3{-}4521.5 \, \text{MeV} \)~\cite{LHCb:2024smc}, differing significantly from the 2011 results~\cite{LHCb:2021uow}. Its partner state is likely \( X_{c1}(4685) \), with a lower mass limit of \( 4667 \, \text{MeV} \) at the \( D_{s1}^{+}D_{s}^{*-} \) threshold, differing by no more than 20 MeV~\cite{ParticleDataGroup:2022pth}.

\textbf{3.~Hidden-charm pentaquark molecules:}
Similar to the limited observed \( Z_{cs} \) states, only two \( P_{cs} \) baryons, \( P_{cs}(4338) \) and \( P_{cs}(4459) \), have been detected experimentally~\cite{LHCb:2020jpq,LHCb:2022ogu}. These are interpreted as \( \bar{D}\Xi_{c} \) and \( \bar{D}^{*}\Xi_{c} \) molecules.  Heavy-quark spin symmetry predicts
two \( \bar{D}^{*}\Xi_{c} \) molecular states~\cite{Peng:2020hql}: a spin \( 3/2^{-} \) state matching \( P_{cs}(4459) \) and an unobserved spin
\( 1/2^{-} \) state \footnote{Ref.~\cite{Zhu:2022wpi} suggests through fitting experimental data that the spin-parity of \( P_{cs}(4459) \) is
\( 1/2^{-} \), while the \( 3/2^{-} \) state has not yet been experimentally discovered}.  We predict this \( 1/2^{-} \) state
appear as a distinct peak in the \( \pi^{0} P_{cs}(4338) \) invariant mass spectrum at \( M = 4477.97 - 4478.2 \) MeV.
Additionally, we suggest searching for two more \( P_{cs} \) molecular states:
the \( \Xi_{c}' \bar{D} \) state (\( M = 4447.86 - 4448.9 \) MeV) and \( \Xi_{c}^{*} \bar{D} \) state (\( M = 4514.76 - 4515.87 \) MeV), associated
with \( \delta(\Xi_{c}^{'+}  \Xi_{c}^+ \gamma) = 110.51 \) MeV and \( \delta(\Xi_{c}^{*+} \Xi_{c}^+ \pi^{0}) = 42.42 \) MeV, respectively.
The existence of these \( P_{cs} \) states has prompted extensive theoretical studies, such as two recent work~\cite{Zhu:2022wpi,Li:2024jlq}.

\textbf{4.~Singly-Charm tetraquark molecules:}
Using the molecular state information of \( X(2900) \) with \( K^{*+}D^{*+} \) as input, we predict the possible existence of a \( D_1^{+}K^{*+} \)
molecular state. Its mass is \( M = 3313.77 - 3323.89 \) MeV, taking \( \delta(D_1(2420)^{+}D^{*+}\pi^0) = 280.86 \) MeV.  We believe that there is
no \( K_1D^{*} \) molecular state due to the large difference in \( \delta(K_1K^{*}\pi) \), with a mass difference greater than 20 MeV. However,
we do not rule out the existence of molecular state structures below the \( K_1D^{*} \) threshold.

\textbf{5.~Hidden-bottom tetraquark molecules:}
According to HQFS, a molecular state containing a \( c\bar{c} \) pair implies a corresponding \( b\bar{b} \) state, and vice versa. However, only two molecular states, \( Z_b(10610) \) and \( Z_b(10650) \)~\cite{Belle:2013urd,Belle:2011aa}, have been identified as \( B\bar{B}^{*} \) and \( B^{*}\bar{B}^{*} \), respectively, as flavor partners of \( X(3872) \) and \( X(4020/4025) \)~\cite{Guo:2013sya}. This scarcity challenges HQFS. We propose experiments to search for the invariant mass of the \( \gamma Z_b(10610) \) final state, where a \( B^{*+}\bar{B}^{*-} \) molecular state with mass \( M = 10655.5 - 10655.6 \) MeV may exist. This state, predicted as the HQSS partner of \( Z_b(10610) \), would support the existence of \( X(4014) \), the HQSS partner of \( X(3872) \). Additionally, the Belle II collaboration found strong evidence for a new molecular state just above the \( \bar{B}^{*}B^{*} \) threshold (2-5 MeV)~\cite{Belle-II:2024niz}.

We propose experiments to search for the \( B_1(5721) \bar{B} \) molecular state, which may appear as a triangle singularity peak in the \( \pi^{0} Z_b(10610) \) invariant mass spectrum, with a mass range of \( M = 11005.4 - 11011.8 \) MeV, near the experimentally observed \( \Upsilon(11020) \) at \( M = 11000 \pm 4 \) MeV~\cite{ParticleDataGroup:2022pth}. Using the lower mass limit for \( B_1(5721) \), the predicted mass range for the molecular state is \( M = 11002.7 - 11009.0 \) MeV, consistent with \( \Upsilon(11020) \), suggesting \( \Upsilon(11020) \) may be a \( B_1(5721) \bar{B} \) molecular state. Additionally, the partner state \( B_1(5721) \bar{B}^{*} \) is predicted to have a mass range of \( M = 11051.7 - 11058.1 \) MeV in the \( \pi^0 Z_b(10650)^{+} \) final state. We also suggest searching for the \( B_{s1}(5828)^{0} \bar{B}^{*0} \) molecular state in the \( K^{-} Z_b(10650)^{+} \) final state with mass \( M = 11154.4 - 11154.9 \) MeV.

\textbf{6.~Three charm/bottom tetraquark molecules:}
An intriguing phenomenon is the absence of hadrons with three charm quarks, despite the detection of singly, doubly, and fully charmed tetraquarks such as \( T_{cs}^-(2900) \), \( T_{cc}^+(3875) \), and \( X(6900) \)~\cite{ParticleDataGroup:2022pth}. We suggest experiments search for such hadrons in the \( D^0X(3872) \) decay. Given \( \delta(\psi(3770)\bar{D}^0D^0) = 44.02 \) MeV and the dominance of \( \psi(3770) \) decaying into \( \bar{D}^0D^0 \)~\cite{ParticleDataGroup:2022pth}, a tetraquark state \( X_{cc\bar{c}} \) with \( c\bar{c}c\bar{q} \) and mass \( M = 5783.96-5799.42 \) MeV could produce a triangular singularity peak in the \( X_{cc\bar{c}} \to \psi(3770)D^{*} \to DX(3872) \) reaction, detectable experimentally. A similar partner state \( X_{bb\bar{b}} \), formed by \( \Upsilon(10885)B^{*+} \), can be explored via the \( B^{*0}Z_b(10650)^{+} \) final state, though its mass variation exceeds 20 MeV, with the maximum close to the value in Eq.~\ref{eq2}.

\begin{table}[http!]
\centering
\caption{This lists the above-threshold molecular states we recommend for experimental searches. Here, $\delta{}M$ is the maximum mass difference from the threshold (in MeV).
The $\surd$ symbol indicates molecular states observed experimentally but needing further analysis, while $?$ denotes states that cannot be explained by the current triangular singularity mechanism.
}\label{tab-2}
\setlength{\tabcolsep}{1.8mm}{
\begin{tabular}{ccc|ccc}
\hline\hline
 Molecule                         & $\delta{}M$      &   Exp.      & Molecule                 &$\delta{}M$  & Exp.                 \\ \hline
 $D^{*0}\bar{D}^{*0}$             & $2.70$           &   $X(4014)$ & $D_1\bar{D}$             & $16.89$     & $Y(4230)$            \\
 $\psi(3770)\bar{D}^{*0}$         & $15.46$          &             & $D_{s1}^{+}D^{*+}$       & $3.06$      &                       \\
 $D_s^{*-}D^{*+}$                 & $2.56$           &   $Z_{cs}(4123)$          & $D^{*}DK$                & $12.00$     &                       \\
 $ \Xi_{c}^{'} \bar{D}$           & $1.04$           &             & $\Xi_{c}^{*} \bar{D}$    & $1.11$      &                       \\
 $ B^{*+}B^{*-}$                  & $0.10$           &   $\surd$   & $B_1 \bar{B}$            & $6.40$      &   $\Upsilon(11020)$   \\
 $B_{s1}^{0} \bar{B}^{*0}$        & $0.50$           &             &$D_1\bar{D}^{*}$           & $17.59$    &   $Z(4430)$         \\
 $D_{s1}^{+}D^{-}$                & $2.94$           &             & $\bar{D}^{*}\Xi_{c}$      & $0.23$                             \\
 $D_1^{+}K^{*+}$                  & $10.12$          &             & $B_1 \bar{B}^{*}$         & $6.40$                               \\
 $D_{s1}^{+}D_{s}^{-}$            & $3.03$           &   $X_{c0}(4500)$          & $D_{s1}^{+}D_{s}^{*-}$         & $19.69$         & $X_{c1}(4685)$ $?$                      \\
\hline \hline
\end{tabular}}
\end{table}
\textit{Summary}-- Given the current experimental challenges in detecting molecular states predicted to be below the theoretical thresholds, we suggest extending the concept of molecular states to include those that may truly exist just above the thresholds, but are very close to them.  This potential has often been overlooked, despite the identification of candidates above these thresholds, such as \( T^{++}_{c\bar{s}0}(2900) \), \( X(3872) \), \( Z_{cs}(3985) \), \( X(4020/4025) \), \( Z_b(10610) \), \( Z_b(10650) \), and \( P_{cs}(4338) \).  Using the information about
these molecular states, we propose using the prominent peaks from the triangle singularity mechanism as experimental signals to search for new molecular states above threshold, particularly heavy quark states to test heavy quark symmetry.

 Our findings are summarized in Tab.~\ref{tab-2}. We suggest experiments on $\gamma X(3872)$ and $\gamma Z_{cs}(3985)$ final states to confirm the $X(4014)$ and $Z_{cs}(4123)$, predicted to contain $D^{*}\bar{D}^{*}$ and $D^{*}\bar{D}^{*}_s$ components, respectively. We also propose that the states $Y(4320)$, $Z(4430)$, $X_{c0}(4500)$, and $\Upsilon(11020)$ correspond to molecular states with $D_1\bar{D}$, $D_1\bar{D}^{*}$, \( D_{s1}^{+}D_{s}^{-} \), and $B_{1}(5721)\bar{B}$, respectively, and recommend using the triangular singularity mechanism for their identification. Further investigation is needed to confirm whether \( X_{c1}(4685) \) can be the partner state of \( X_{c0}(4500) \), with \( D_{s1}^{+}D_{s}^{*-} \) components, as current models cannot predict its mass. Lastly, we recommend that BES II, III, the LHC Collaboration et al investigate eleven additional predicted heavy quark molecular states, particularly the \( B^{*+}B^{*-} \) states, for which experimental evidence already exists.
The potential discovery of these states would significantly advance our understanding of heavy quark symmetry and illuminate the mechanisms governing the formation of molecular states above threshold arising from hadronic interactions.

\section*{Acknowledgments}

This work is supported by National Key R\&{}D Program of China No.2024YFE0109800 and 2024YFE0109802.

\end{document}